\documentclass[fleqn,usenatbib]{mnras}

\usepackage{newtxtext,newtxmath}

\usepackage[T1]{fontenc}

\DeclareRobustCommand{\VAN}[3]{#2}
\let\VANthebibliography\thebibliography
\def\thebibliography{\DeclareRobustCommand{\VAN}[3]{##3}\VANthebibliography}

\usepackage{graphicx}	
\usepackage{amsmath}	
\usepackage[dvipsnames]{xcolor}

\title[Fixed vs live binary orbits]{ The importance of live binary evolution in numerical simulations of binaries embedded in circumbinary discs}

\author[Franchini et al.]{
Alessia Franchini,$^{1,2}$\thanks{E-mail: alessia.franchini@unimib.it}
Alessandro Lupi,$^{1,2,3}$
Alberto Sesana,$^{1,2}$
Zoltan Haiman$^{4}$
\\
$^{1}$ Dipartimento di Fisica ``G. Occhialini", Universit\'a degli Studi di Milano-Bicocca, Piazza della Scienza 3, 20126 Milan, Italy\\
$^{2}$ INFN, Sezione di Milano-Bicocca, Piazza della Scienza 3, 20126 Milano, Italy\\
$^{3}$DiSAT, Universit\`a degli Studi dell'Insubria, via Valleggio 11, I-22100 Como, Italy\\
$^{4}$ Department of Astronomy, Columbia University, New York, NY 10027, USA
}

\begin{document}
\label{firstpage}
\pagerange{\pageref{firstpage}--\pageref{lastpage}}
\maketitle

\begin{abstract}
The shrinking of a binary orbit driven by the interaction with a gaseous circumbinary disc, initially advocated as a potential way to catalyze the binary merger, has been recently debated in the case of geometrically thick (i.e. with $H/R\gtrsim 0.1$) discs. However, 
a clear consensus is still missing mainly owing to numerical limitations, such as fixed orbit binaries or lack of resolution inside the cavity carved by the binary in its circumbinary disc.
In this work, we asses the importance of evolving the binary orbit by means of hydrodynamic simulations performed with the code {\sc gizmo} in meshless-finite-mass mode. In order to model the interaction between equal mass circular binaries and their locally isothermal circumbinary discs, we enforce hyper-Lagrangian resolution inside the cavity.
We find that fixing the binary orbit ultimately leads to an overestimate of the gravitational torque that the gas exerts on the binary, and in an underestimate of the torque due to the accretion of material onto the binary components.  
Furthermore, we find that the modulation of the accretion rate on the binary orbital period is strongly suppressed in the fixed orbit simulation, while it is clearly present in the live binary simulations. This has potential implications for the prediction of the observable periodicities in  massive black hole binary candidates.

\end{abstract}




\section{Introduction} 
\label{sec:intro}

Circumbinary discs are expected to surround both stellar binaries \citep[e.g.][]{chiang2004,kennedy2019} and massive black hole binaries (MBHBs).  Although we still lack unambiguous observational evidence of the latter, any cloud of gas approaching the binary on a non zero angular momentum orbit is expected to form a circumbinary disk due to the conservation of angular momentum \citep{Pringle1981,lodato2008}.

Once formed, a coplanar circumbinary disc can extend down to a few times the binary separation \citep{artymowicz1994}. The circumbinary disc orbits resonate with the binary orbit at discrete locations (Lindblad resonances), leading to the exchange of angular momentum between the disc and the binary \citep{lynden-bell1972,Lin1986}. The magnitude of the resonant torques depends on the binary potential, i.e. its mass ratio and eccentricity, and is proportional to the disc surface density at the resonance locations \citep{goldreich1979}. Therefore, the amount of angular momentum transferred from the binary to the disc at the resonances depends on the disc properties as well. 
In the equal mass binary regime however the binary evolution is determined by the properties of the discs that form around each binary component from the material that leaks into the cavity. In particular, the gravitational torque exerted by these circumstellar discs can be positive and, in a specific region of the parameter space, overcome the negative gravitational torque of the circumbinary disc.

Recent 2D hydrodynamic simulations have suggested that the exchange of angular momentum between the disc and the binary can also lead to binary expansion, contradicting previous predictions \citep{Munoz2019,Munoz2020}. In particular, for locally isothermal discs, several  2D with fixed binary orbits and 3D with live binaries works have argued that the disc aspect ratio determines whether the binary transfers angular momentum to the disc or viceversa 
\citep{tiede2020,heathnixon2020}. However, using very high resolution 3D hydrodynamical simulations of live binaries, \cite{Franchini2022} found this process not to be simply regulated by the disc temperature but also by the disc viscosity. 
The picture becomes even more complicated if we consider the specific numerical limitations and assumptions of the different approaches employed.
In this respect, the main difference between previous works is that all the employed grid codes both in 2D and 3D \citep{Moody2019} assumed the binary orbit to remain fixed with time whereas 3D particle codes evolved by construction the binary orbit together with the gas particles, conserving linear and angular momentum.

In this work, we want to assess the importance of evolving the binary orbit and to understand whether not evolving the orbit for initially circular equal mass binaries may alter the outcome of binary-disc interactions.
We do not specifically focus on either stellar or massive black hole binaries as our simulations are scale free. We, however, note that the disc aspect ratios employed in this work are more compatible with the ones expected in protoplanetary discs rather than cold AGN-like discs around massive black hole binaries \citep[][]{Collin1990}.

We use the code {\sc gizmo} \citep{Hopkins2015} in its 3D mesh-less finite mass (MFM) mode, coupled with adaptive particle-splitting for numerical refinement of the gas dynamics inside the disc cavity (for details see \citet{Franchini2022}), to investigate the difference in terms of gas distribution between simulations that keep the binary orbit fixed in time and simulations of live binaries.

In Section \ref{sec:sim}, we describe the numerical method, and present the results of our simulations in Section \ref{sec:results}. We discuss the relevance of the results in Section \ref{sec:discussion} and finally draw our conclusions in Section \ref{sec:concl}.

\section{Numerical setup} 
\label{sec:sim}

The initial conditions for this work consist of a binary surrounded by a circumbinary gaseous disc. The 3D distribution of the $10^6$ equal mass gas particles \footnote{ We note that since we have performed convergence tests in our previous work \citep{Franchini2022}, we do not repeat those tests in the fixed binary orbit case as we have reached a high enough resolution for the torques value to converge.} sampling the disc and the initial orbit of the two sink particles of the binary are generated using the SPH code {\sc phantom} \citep{price2017}. 
The equal mass binary has an initial mass $M=M_1+M_2=1$, a separation $a=1$ and an orbital period $P_{\rm b}=2\pi(a^3/GM)^{1/2}$. The circumbinary disc initially extends from $R_{\rm in} =2a$ to $R_{\rm out}=10a$, with a fixed aspect ratio $H/R=0.1$. The gas is described by a locally isothermal  equation of state with the sound speed $c_{\rm s}$ defined by Eq.~(4) in \cite{farris2014}, and a surface density profile scaling as $\Sigma \propto R^{-3/2}$, normalised to get a total mass $M_{\rm disc}=0.1M$. 
 The physical units used in the simulations correspond to a solar mass binary with a semi-major axis of 1 AU. However our simulations are naturally scale free and can be applied in different regimes of physical scales, from stellar binaries to massive black hole binaries.

The simulations are performed with the MFM method in {\sc gizmo}.
We here also include the effect of gas viscosity entering the Navier-Stokes fluid equations as described in \citet{hopkins2016}, assuming a shear viscosity in the disc $\nu=\alpha c_{\rm s}H$, parametrised using a viscosity parameter $\alpha=0.1$ \citep{ss1973}, and no bulk viscosity. 

In order to increase the resolution when and where necessary, without globally slowing down our simulations, we employ an on-the-fly adaptive particle-splitting approach, which is similar in spirit to the adaptive mesh refinement of grid-based codes. Such technique represents a natural generalisation of the refinement/de-refinement scheme in {\sc arepo} \citep{Springel2010} and {\sc gizmo} \citep{Hopkins2015} to maintain an almost constant mass per cell during the simulations when a finite-volume scheme is employed.
In this work, we exploit the particle splitting algorithm in {\sc gizmo}, splitting gas particles that enter a sphere of radius $r_{\rm ref}=4a$ centred on the centre of mass of the binary-disc system. The refinement scheme is the same as the one adopted in \cite{Franchini2022} with a maximum refinement of a factor $32$. 

Every time a gas particle approaches one of the sinks, entering its sink radius $r_{\rm sink}$, we flag it as eligible to be accreted. In the live binary case, conservation of mass, and linear and angular momentum are ensured during each accretion event, in the same way it is done in the {\sc phantom} code \citep{bate1995}.

In this work, we neglect the disc self-gravity (see \cite{franchini2021} for self-gravity treatment) and use the version of the code described in details in \cite{Franchini2022}.  The reason for this choice is that the inclusion of the disc self-gravity would make the evolution more complex, as other processes would come into play (e.g. cooling, gravitational instabilities), and we wanted to perform this comparison in the simplest possible scenario. In addition, self-gravity can be neglected for sufficiently compact binaries, for which the disc mass is negligible
(see \citealt{franchini2021} for a discussion). The extension of this work to the self-gravity regime will be performed in a future study.

\subsection{Fixed binary setup}

We let the live binary evolve for $\sim 4000$ orbits and we then take the $1000$th orbit as initial condition for the fixed binary orbit simulation since, by this time, the disc is in a quasi-steady state. Starting from a live binary snapshot, we fix the binary orbit and keep tracking the evolution of the gas for $700$ orbits. We remove the particles that would accrete onto the binary following the same approach of the live binary simulation but without changing the properties of the binary. This implies that angular momentum is not conserved.
At this time, the binary has total mass $M=1.04$ and separation $a=0.97$.

Note that, even though the binary orbit is kept fixed, we keep track of the accreted mass and angular momentum onto the binary throughout the simulation.

 The choice we made to start from a relaxed snapshot of the live binary simulation is motivated by the fact that the system experiences a transient behaviour within the first few hundreds orbits, as shown in \cite{Franchini2022}, as the disc is not initially set at perfect equilibrium. We therefore opted to start when the system had already approached a quasi-steady state, and we could rely on the subsequent evolution without risking the initial conditions affecting our conclusions. 

\begin{figure}
    \centering
    \includegraphics[width=\columnwidth]{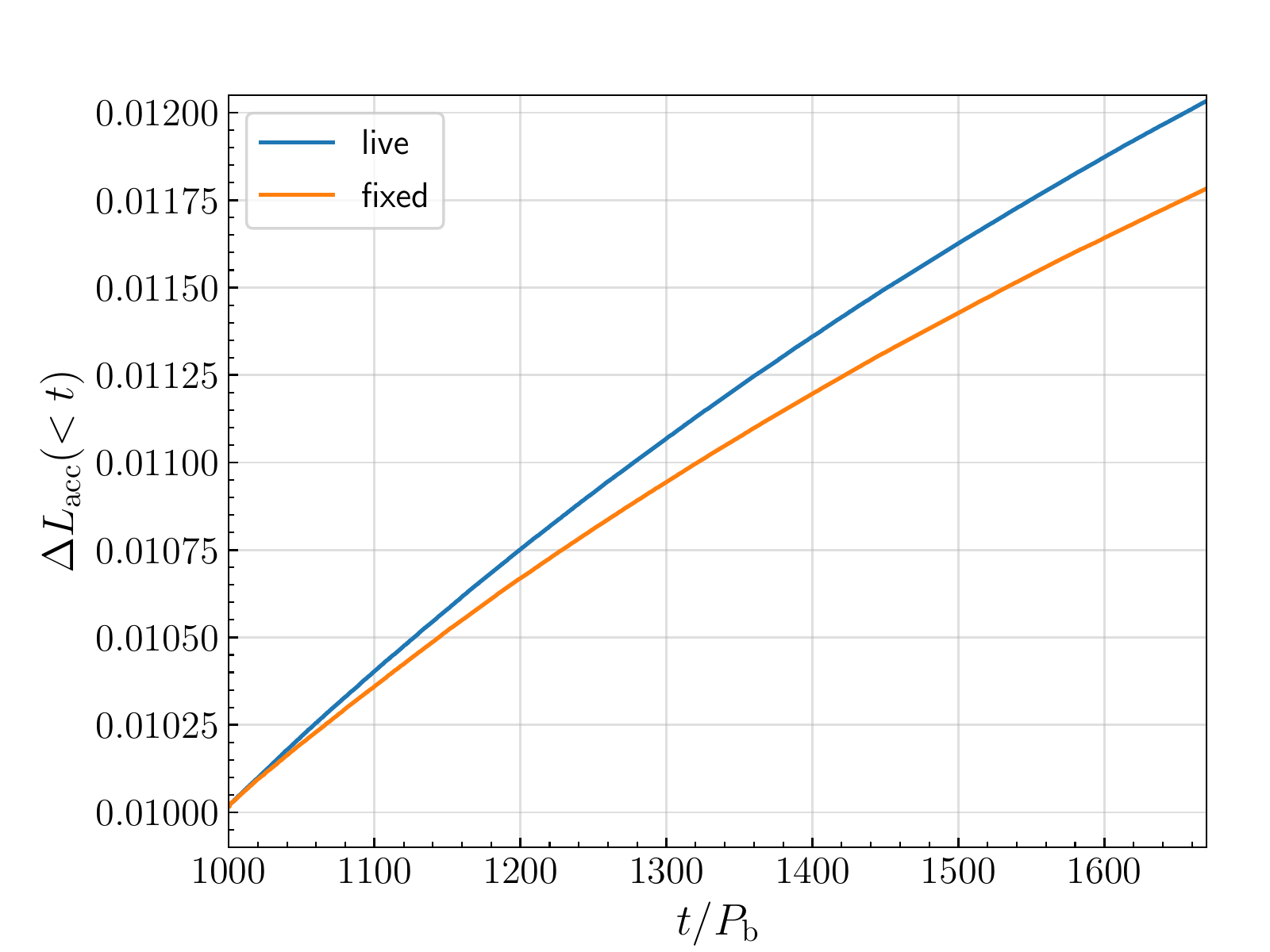}
    \caption{Cumulative accretion torque (orbital angular momentum + spin) contribution to the binary angular momentum change as a function of time in units of the binary orbital period.} 
    \label{fig:deltal}
\end{figure}

\begin{figure}
    \centering
    \includegraphics[width=\columnwidth]{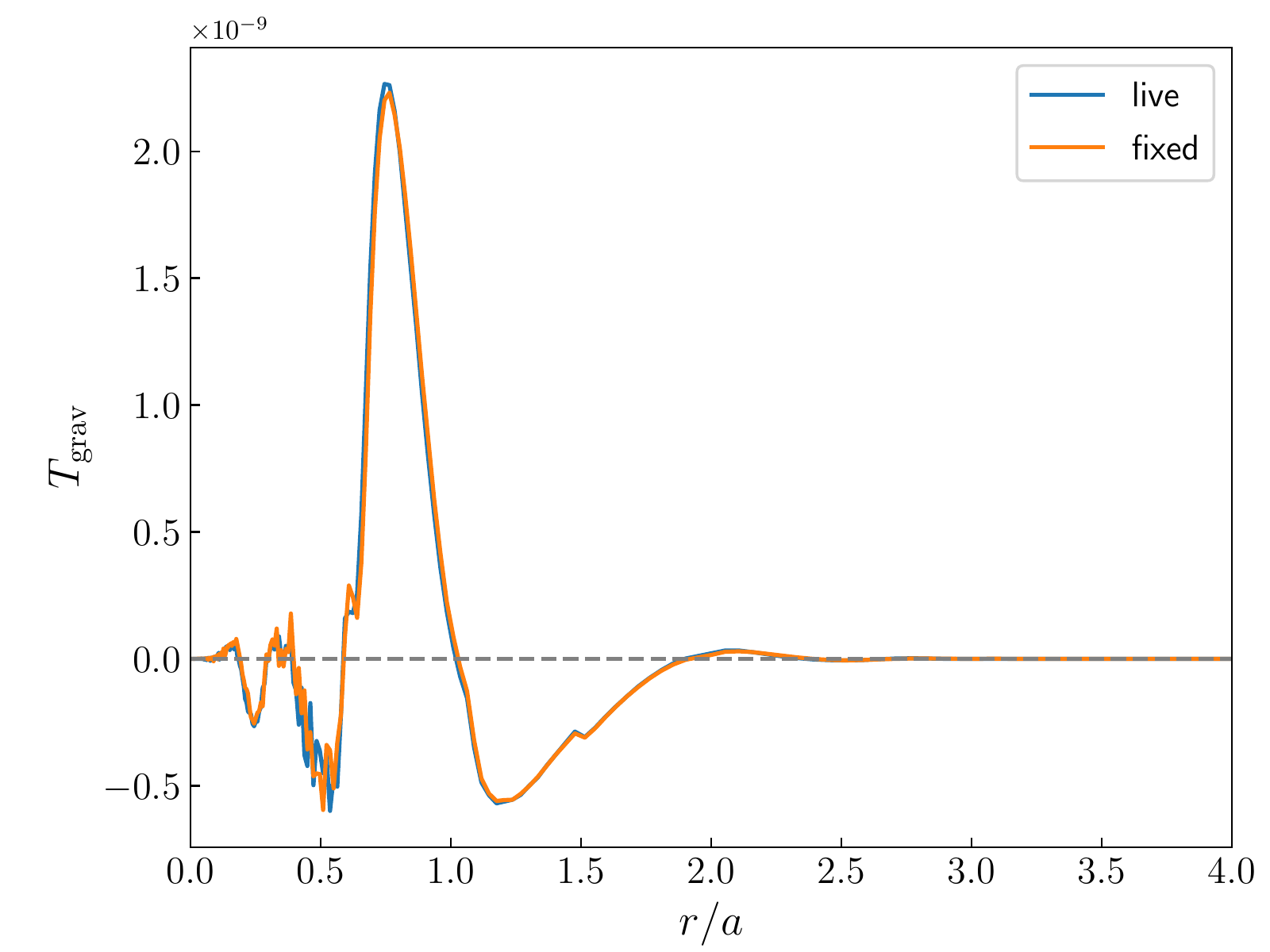}
    \caption{Torque profile integrated over the azimuthal angle inside $4a$, averaged over $1600-1700\,P_{\rm b}$ . }
    \label{fig:torqueprof}
\end{figure}

\begin{figure*}
    \includegraphics[width=\textwidth]{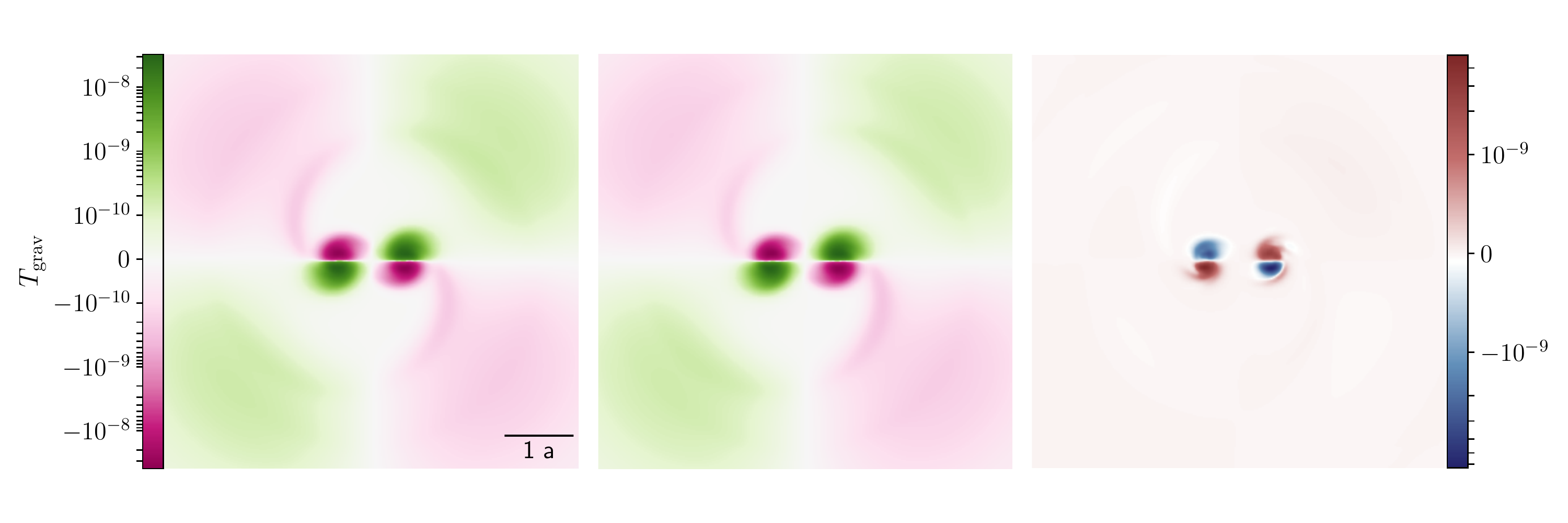}
    \caption{Total gravitational torque distribution inside the cavity averaged over the last 100 orbits (i.e. $1600-1700\,P_{\rm b}$). Left panel: live orbit binary. Middle panel: fixed binary. Right panel: difference between the live and fixed binary orbit map. }
    \label{fig:grav_maps}
\end{figure*}

\section{Results} \label{sec:results}


We here describe the results of our analysis comparing the torques in the live and fixed orbit binary simulations.

\subsection{Gas accretion and gravitational forces}

The gravitational torque that the distribution of gas particles exerts on the binary is 
\begin{equation}
    {\bf T}_{\rm G} = \sum_{i=1}^{N}{\bf r}_1\times \frac{GM_1m_{\rm i}({\bf r}_{\rm i}-{\bf r}_1)}{|{\bf r}_{\rm i}-{\bf r}_1|^3} + \sum_{i=1}^{N}{\bf r}_2\times \frac{GM_2m_{\rm i}({\bf r}_{\rm i}-{\bf r}_2)}{|{\bf r}_{\rm i}-{\bf r}_2|^3}
    \label{eq:gravt}
\end{equation}
where $m_{\rm i}$ is the particle mass and $r_1,r_2$ are the sinks positions with respect to the binary centre of mass.
Since this quantity is subject to fluctuations on very short timescales, we calculate the average value of the gravitational torque for both the live and fixed binary simulations over the $700$ orbits that we are considering. 
We find that in the fixed orbit case, the time-averaged gravitational torque value is higher by $\sim$ 5\% compared to the live binary case.

When a gas particle is accreted, the binary angular momentum changes by

\begin{equation}
 d{\bf L}_{\rm acc} = {\bf r}_{\rm i} \times m_{\rm i}{\bf v}_{\rm i} - \frac{m_{\rm i}M_{\rm k}}{(m_{\rm i}+M_{\rm k})}\left[({\bf r}_{\rm i}-{\bf r}_{\rm k})\times ({\bf v}_{\rm i}-{\bf v}_{\rm k})\right]
 \label{eq:dLacc}
\end{equation}
where $r_{\rm i},{\bf v}_{\rm i}$ are the position and velocity of the gas particle that is being accreted and $r_{\rm k},{\bf v}_{\rm k}$ are the position and velocity of the sink particle that is accreting. Note that the second term on the r.h.s. contributes to the spin of each binary component.
In the fixed binary orbit simulation we do not allow the mass of the binary to increase owing to accretion but we simply remove the accreted particles from the simulation storing its properties in order to compute $d{\bf L}_{\rm acc}$ a posteriori.

Figure \ref{fig:deltal} shows the cumulative (i.e. sum over time) 
accretion torque as a function of time for the live (blue line) and fixed orbit (orange line) binary.
We can see that the accretion torque is systematically under-estimated by $\sim$9\% over $700\,P_{\rm b}$. 
The rate of angular momentum change due to accretion is slightly steeper in the fixed binary orbit case, therefore we expect the difference in accretion torque between the two cases to become more significant over time.

The difference in the accretion torque is likely due to the difference in the accretion rate between the two simulations, as the velocities of the accreted particles are similar within 1\%, while the accretion rate is 7\% higher in the live case compared to the fixed binary orbit case.

Since we are interested in comparing the entity of the gravitational torque change due to the possibly different gas distribution between the two simulations, we compute the gravitational torque profile inside $r=4a$ averaged over the last 100 orbits.
The results are shown in Figure \ref{fig:torqueprof} where the blue and orange line represent the live and fixed orbit simulation respectively.
The sink particle lies at $r\sim 0.5a$ and the strong positive contribution to the gravitational torque comes from the external part of the circumstellar disc while the stream that feeds this disc exerts a negative gravitational torque on the binary.
In general, the magnitude of the gravitational torque averaged over $100\,P_{\rm b}$ is essentially comparable in the two simulations. We chose to average the torque over the last 100 orbits as by this time the system has reached a quasi-steady state.

\subsection{Gravitational torque maps}

In order to understand better the different torques profiles of Fig. \ref{fig:torqueprof}, we produced maps of the gravitational torque density for both simulations.
Figure \ref{fig:grav_maps} shows the torque density within the cavity carved by the binary in the circumbinary disc. The left (middle) panel represents the fixed (live) binary simulation averaged over the last 100 orbits, i.e. in the interval $t=1600-1700\,P_{\rm b}$. The rightmost panel shows the difference between the live and fixed binary orbit map.
Note that we have rotated the particles in order for them to be in the binary reference frame so that we can do a straightforward comparison between the gas distribution in the two simulations. 

We can see that there are tiny differences in the two maps, mostly in the circumstellar discs. 
In regions where the torque is positive (negative), the difference is positive (negative). This implies that in the live binary orbit case the gravitational torque is stronger with respect to the fixed binary case. This is consistent with having more massive circumstellar discs in the live binary simulation (see below Fig. \ref{fig:minidiscSigma}).  
We can however see that the distribution of negative gravitational torque in the rightmost panel is slightly more asymmetrical with respect to the positive quadrants. Therefore the overall gravitational torque exerted by the circumstellar disc is lower (i.e. less positive) in the live binary orbit simulation.

\subsection{Circumstellar discs density}
\begin{figure}
    \centering
    \includegraphics[width=\columnwidth]{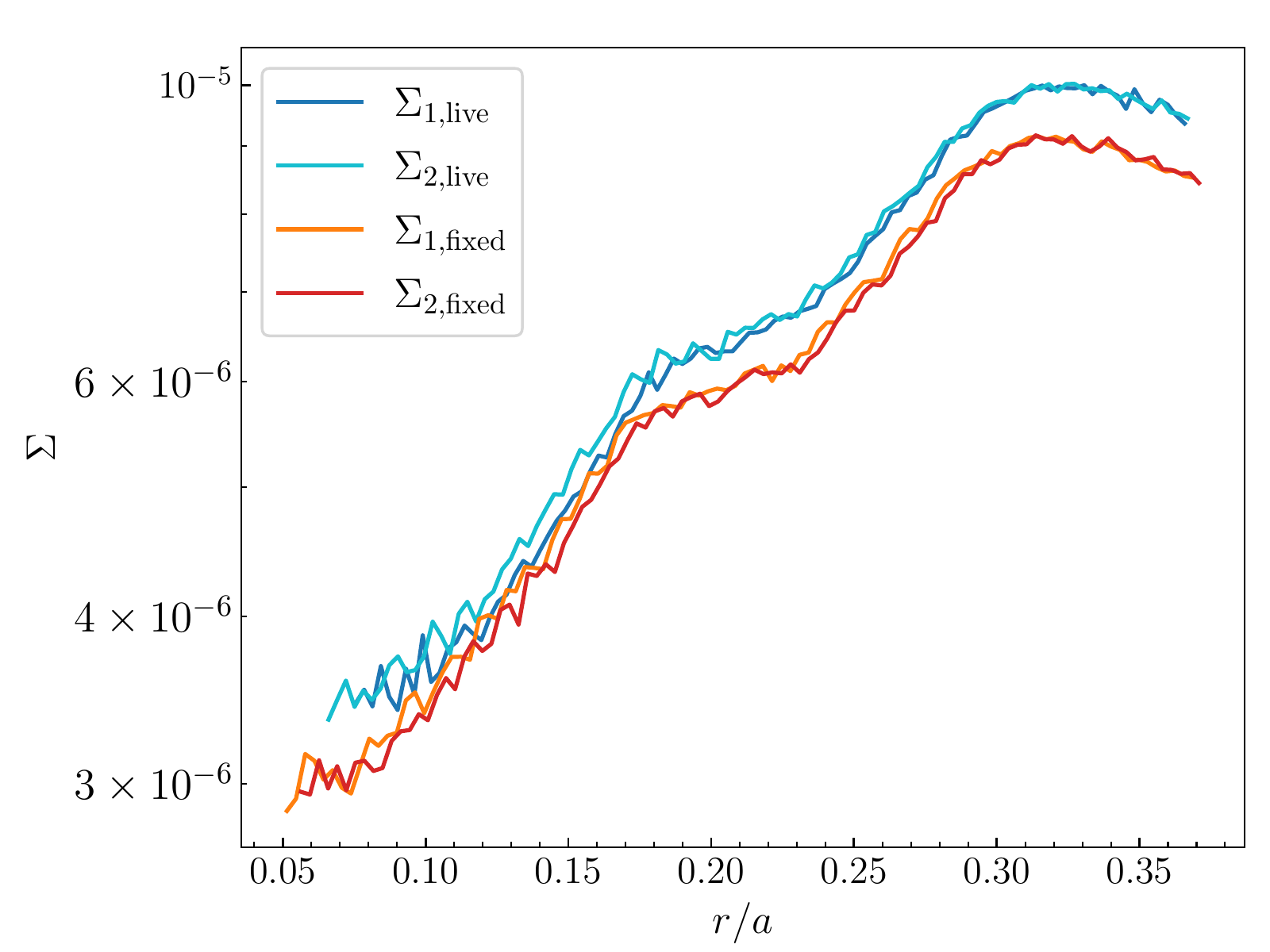}
    \caption{Circumstellar discs surface density profiles in the live (blue and cyan lines) and fixed (orange and red lines) simulations as a function of radius in units of the initial binary separation.}
    \label{fig:minidiscSigma}
\end{figure}
The gas leaks into the cavity through a stream that forms and continuously supplies material to circumstellar discs.
We averaged the surface density profile of these discs, $\Sigma_1$ and $\Sigma_2$, over the last 100 orbits of each simulation and we show the results in Figure \ref{fig:minidiscSigma}.

 The blue and cyan lines show the profiles in the live binary simulation while the orange and red lines show the profiles obtained in the fixed binary orbit simulation.
The circumstellar discs around the binary components are almost perfectly symmetric in both simulations. 
However, their density is slightly lower in the fixed binary orbit simulation.
The quantity of material that is being flung back from the cavity into the disc is slightly larger in the fixed orbit simulation. The binary on a fixed orbit exerts always the same torque on the gas that enters the cavity and therefore we expect the properties of the flung back material to remain the same over time.
The amount of material that is contained in the disc beyond $r=2a$ is indeed slightly larger in the fixed binary orbit simulation over the last 100 orbits.

\subsection{Cavity shape}

The cavity carved by the binary into the circumbinary disc is expected to become eccentric as a result of the interaction with the binary \citep[][]{dangelo2006,dunhill2015,dorazio2016,Ragusa2020}, although to date the responsible mechanisms have not been investigated in details. Even a very small initial spurious non-zero binary orbital eccentricity can trigger the eccentric Lindblad resonances which drive both binary and disc eccentricity growth \citep{TeyssandierOgilvie2016}.
Previous studies have found that, if there is material located at either an outer circular Lindblad resonance (at $R_{\rm L}>a$) or at a non-co-orbital eccentric Lindblad resonance (at $R_{\rm L} \neq a$), the disc becomes eccentric.
\cite{Macfadyen2008} showed that the outer circular Lindblad resonance located at the 3:2 binary-to-disc frequency commensurability, i.e. at $1.59a$ is strong enough to trigger eccentricity growth in equal mass binaries.

We calculate the disc eccentricity profile, averaging it over the last 100 binary orbits, finding that the cavity  
is slightly more eccentric in the fixed binary orbit simulation, probably due to the fact that the trajectories of the gas particles remain essentially unvaried if the binary does not change its orbit with time.

We find an exponential growth  of the disc eccentricity in our live binary simulation after the first few tens of binary orbits, despite the very low binary eccentricity (i.e. $e_{\rm b}=3\times10^{-5}$). The material leaking into the cavity indeed reaches  $1.59a$ within the first $\sim 10$ orbits and is therefore able to excite one of the Lindblad resonances thought to be responsible for disc eccentricity growth.

Since we started the fixed binary orbit run from the 1000th orbit of the live binary simulation, the disc in this case starts already eccentric and the evolution of its eccentricity is very similar to the live binary simulation.

\begin{figure}
    \centering
    \includegraphics[width=\columnwidth]{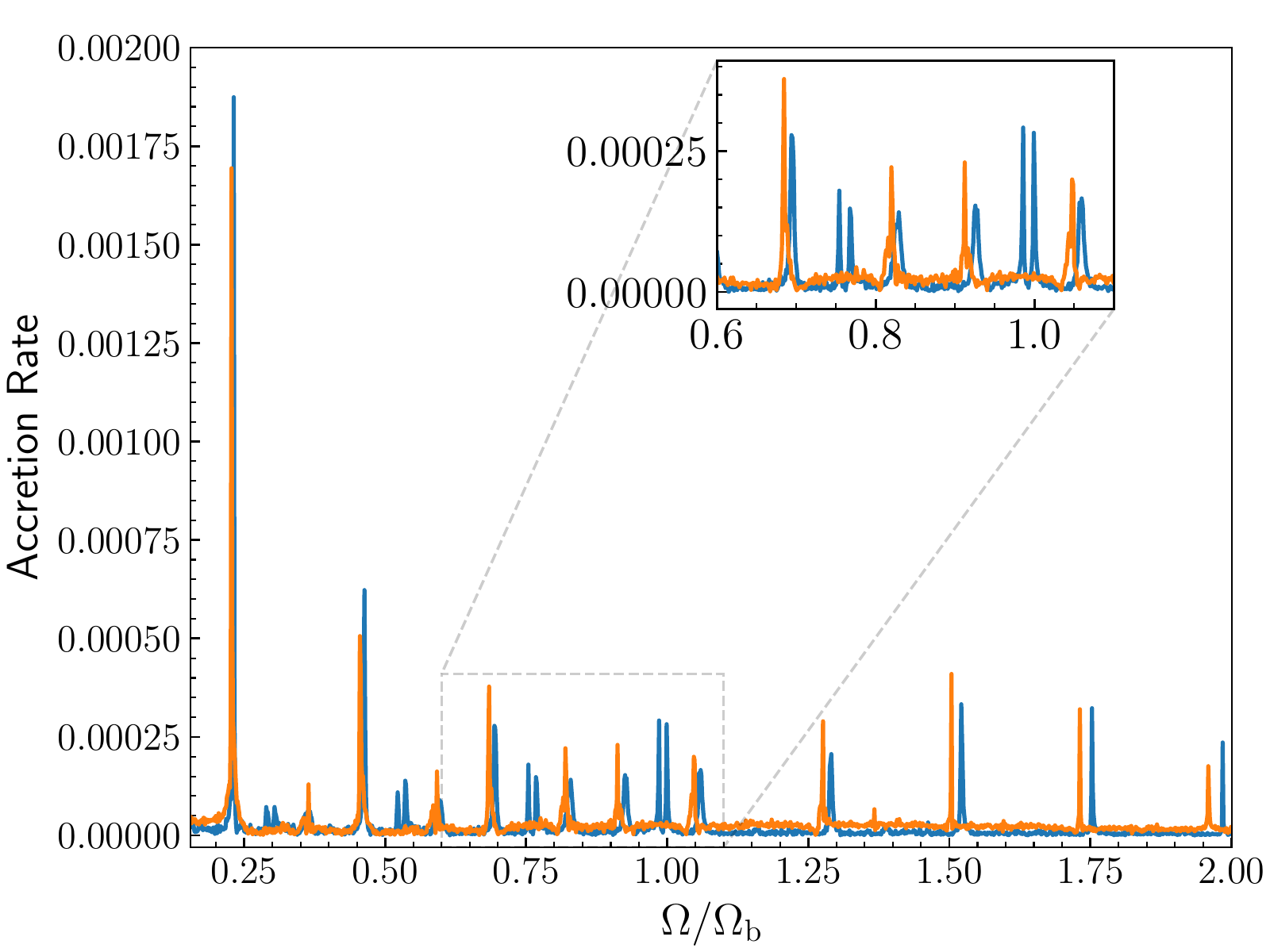}
    \caption{
    Fast Fourier Transform (FFT) of the accretion rate onto the binary for the live (blue line) and fixed (orange line) binary orbit simulation. The initial binary orbital frequency is denoted by $\Omega_{\rm b,0}$.}
    \label{fig:spectrograms}
\end{figure}

\begin{figure}
    \centering
    \includegraphics[width=\columnwidth]{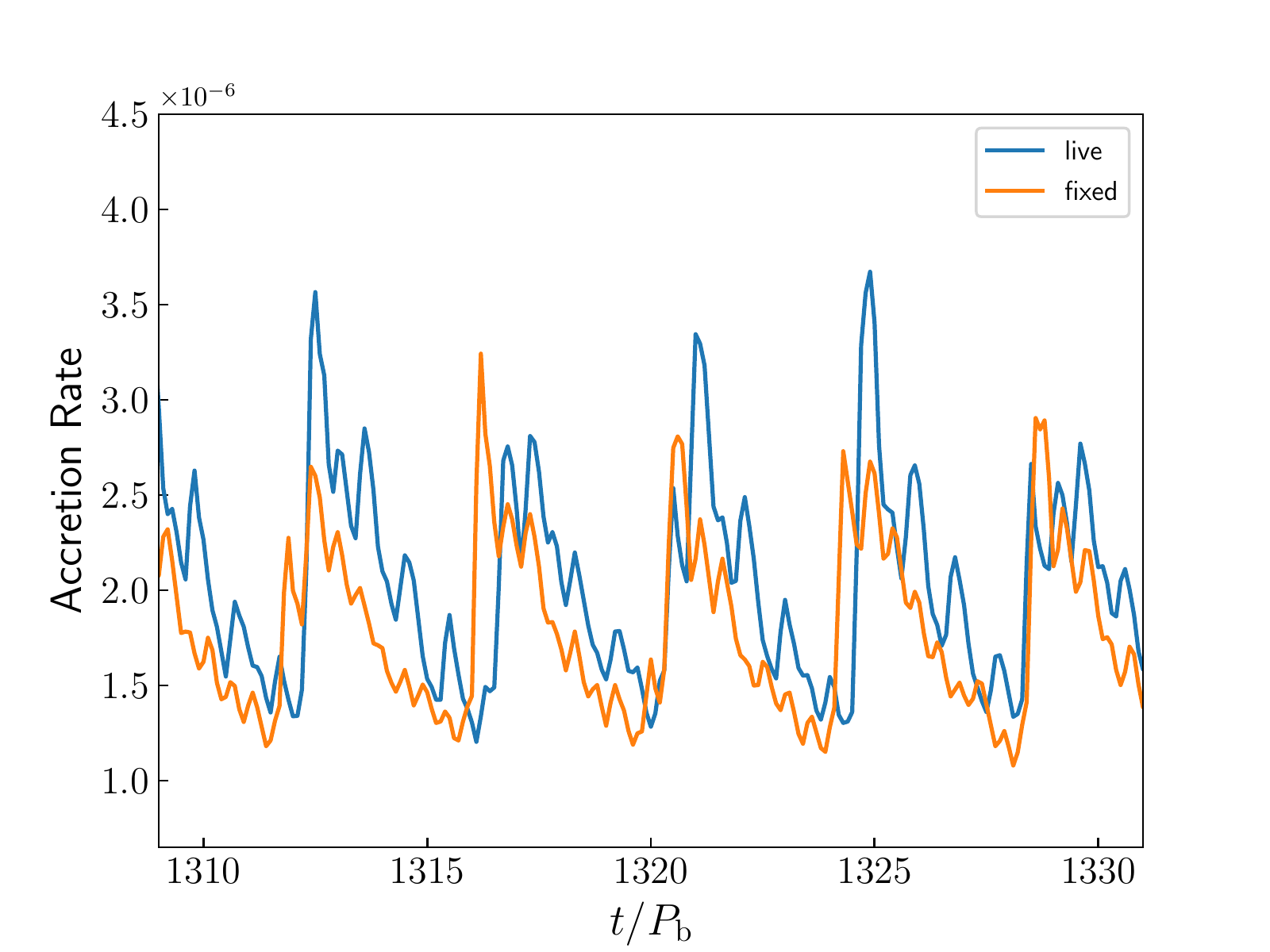}
    \caption{Accretion rate onto the binary for the live and fixed binary simulations, represented by the blue and orange line respectively, as a function of time in units of the binary orbital period.}
    \label{fig:mdot}
\end{figure}

\subsection{Accretion rate modulation}

Since by $1000\,P{\rm b}$ the disc in the live binary simulation has reached a quasi-steady state, we can analyze the variability of the accretion rate onto the binary $\dot{M}$.
This study is important in order to understand which periodicities we can expect to observe from these systems. 

We calculated the Fast Fourier Transform (FFT) of the accretion rate on the binary in the two simulations. The result is shown in Figure \ref{fig:spectrograms} where the blue and orange line represent the live and fixed binary respectively.
The most striking difference between the fixed and live binary simulation is that the modulation on the binary orbital period is present only in the latter, i.e. the 
peak at $\Omega/\Omega_{\rm b,0}=1$ is missing in the fixed orbit simulation.
The lack of this feature has also been observed in \cite{farris2014} and more recently in \cite{Munoz2020} for equal mass circular binaries similar to the ones we are treating here. 
Note that the two adjacent peaks at $\Omega/\Omega_{\rm b,0}=1$ correspond to the natural variation of the binary orbital frequency in the live binary simulation.

Figure \ref{fig:mdot} shows the accretion rate onto the binary calculated directly from the simulation of both the live (blue line) and fixed (orange line) binary. The periodicity on the orbital period of the binary is more pronounced in the live binary case, i.e. the oscillations amplitude is larger, while it seems to be swamped by the lump timescale in the fixed binary simulation. 

\begin{figure}
    \centering
    \includegraphics[width=\columnwidth]{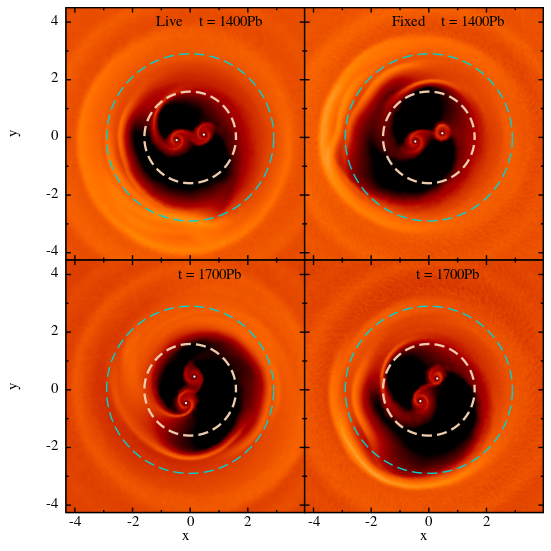}
    \caption{Column density maps of the live and fixed binary orbit simulations at $t=1400\,P_{\rm b}$ (upper panels) and $t=1700\,P_{\rm b}$ (lower panels). The white and cyan dashed lines are located at $1.59a$ (corresponding to the 3:2 resonance) and $2.9a$ (corresponding to the lump) respectively. }
    \label{fig:resonances}
\end{figure}

The periodograms of both simulations however show the same very strong amplitude modulation at $\Omega/\Omega_{\rm b,0}\simeq0.2$ which has been argued to represent the periodicity due to the lump that forms at the disc cavity edge.
The lump at the cavity edge is denser in the live binary orbit simulation by a few percent, whereas the cavity is slightly larger in the fixed binary orbit simulation (see Figure \ref{fig:resonances}).

Another prominent periodicity that has roughly the same amplitude in both simulations is the one corresponding to $\Omega/\Omega_{\rm b,0}\simeq 0.45$.
This corresponds essentially to the location of the stream of gas that is feeding the circumstellar discs, which is very close to the 3:2 binary-to-disc frequency commensurability.

In order to see how much material is contained at these radii in the two simulations, we show in Figure \ref{fig:resonances} the column density maps of the accretion discs at different times.
The upper and lower panels correspond to $t=1400\,P_{\rm b}$ and $t=1700\,P_{\rm b}$ respectively. The cyan dashed circle is located at $r=2.9a$, i.e. the distance that corresponds to the $0.2\,\Omega_{\rm b,0}$ modulation, while the white dashed circle is located at the distance that corresponds to the $0.45\,\Omega_{\rm b,0}$ orbital modulation.

\begin{figure}
    \centering
    \includegraphics[width=\columnwidth]{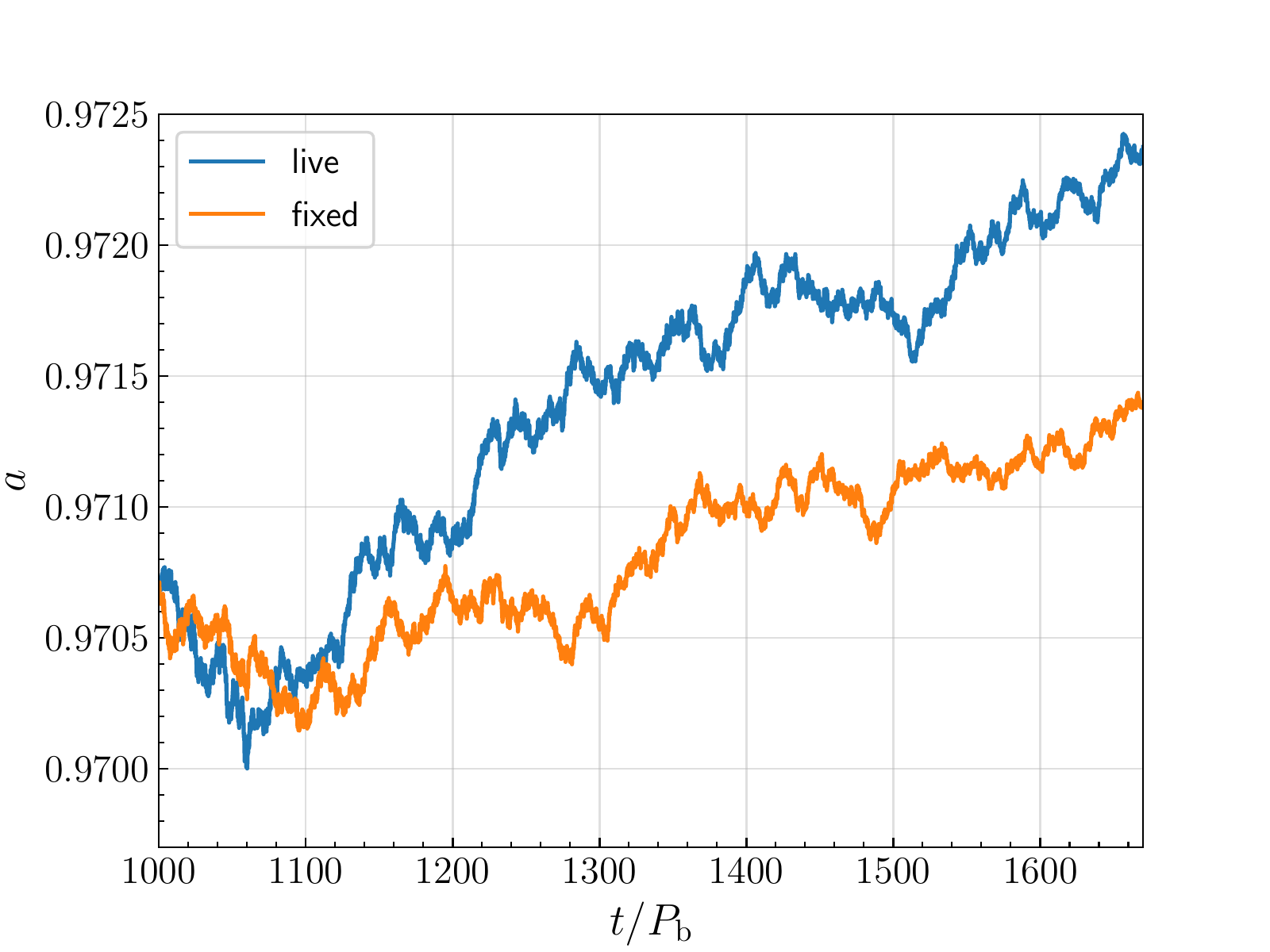}
    \caption{Evolution of the binary semi-major axis in the live (blue line) and fixed (orange line) binary orbit simulation as a function of time in units of the binary orbital period. Note that, in the fixed orbit simulation, we measured the evolution of $a$ analytically using the numerically calculated $\dot{a}$.}
    \label{fig:avst}
\end{figure}

\section{Discussion}
\label{sec:discussion}

Fixing binary orbits in circumbinary discs simulations is a widely adopted strategy, in particular when running 2D (but sometimes also in 3D) simulations using grid codes (either with static or moving mesh grids).
Since in particle codes is much easier to run both live binary and fixed binary orbit simulations, we test whether this approximation can lead to significant differences in terms of gas dynamics and therefore in terms of torques exerted by the gas onto the binary.

We find several differences between the two runs:
\begin{itemize}
    \item Fixing the binary orbit leads to an overestimate of the gravitational torque exerted by the disc onto the binary by about 5\%. Since the balance between the positive torque coming from the circumstellar disc and the negative torque coming from the circumbinary disc is very delicate, a systematic overestimate of one of the two contribution can bias the outcome in terms of binary evolution.
    
    \item Simulations that use fixed binary orbits (in both 2D and 3D) have shown that the accretion contribution to the torque exerted on the binary is always negligible compared to the gravitational torque. However, we here show that fixing the binary orbit does indeed lead to underestimating the accretion torque by 9\% over the 700 orbits we ran. Furthermore, we can see from the upper panel in Figure \ref{fig:deltal} that the difference between the two torques keeps increasing over time and might be even more significant after a few thousands of binary orbits.
    
    \item We find the circumstellar discs density to be slightly higher in the live binary orbit simulation. There is indeed more mass in the circumbinary disc surrounding the fixed orbit binary, possibly because a larger amount of material is flung back in the disc by the binary in this case.
    
    \item The disc eccentricity profile in the two simulations is very similar. 
    We can see from the maps shown in Figure \ref{fig:resonances} that the cavity is slightly larger in the fixed binary orbit simulation.
    This is likely due to the fact that, since the binary properties do not change with time, the amount of material that the binary flings back from inside the cavity into the disc, which is thought to be responsible for the formation of the lump, has always the same properties. 
    
    \item Another important difference between the fixed and live binary orbit simulation is the modulation of the accretion rate onto the binary. While the modulation on the lump timescale is present in both simulations, despite being stronger in the fixed binary orbit run, the modulation of the accretion rate on the binary orbital period is completely missing in the fixed binary orbit case. As expected, this is consistent with previous studies employing fixed binary orbit simulations \citep[][]{farris2014,Munoz2020}. Nonetheless, our finding that in realistic binaries (where the orbital parameters evolve over time) this modulation should be observed also for equal mass binaries is a fundamental result for, e.g., following possible detections of massive black hole binaries through electromagnetic emission \cite{Charisi2016}.
    
    \item Since we ultimately would like to understand the impact that fixing the binary orbit has on the binary evolution, we computed the semi-major axis for both runs using Eq. 2 in \cite{Franchini2022} and assuming $\dot{e}=0$ in the fixed binary simulation as the evolution of the binary eccentricity is negligible in the live binary run and we start fixing the binary orbit at $t=1000\,P_{\rm b}$.
    The results are shown in Figure \ref{fig:avst} where the blue and orange line show the results for the live and fixed orbit respectively.
    We can see that the semi-major axis in the fixed orbit case increases on a longer timescale compared to the live binary case. 
    This is consistent with the fact that the positive contribution of the accretion torque is underestimated in the fixed binary orbit simulation.

\end{itemize}

\section{Conclusions} 
\label{sec:concl}

In this work, we have compared two hydrodynamics simulations performed with \textsc{gizmo} MFM of a circumbinary accretion disc surrounding a live binary and a binary on a fixed orbit.
Although the assumption of fixed binary orbits is very common when exploring binaries embedded in circumbinary discs, we showed that there are several important differences with live binary orbits simulations, meaning that the simplest assumption of fixed orbits can potentially lead to wrong conclusions in terms of binary evolution. We indeed find the $\dot{a}/a$ to be up to a factor $\sim2$ larger, i.e. more positive, in the live binary orbit case, indicating that the binary semi-major axis evolves more rapidly if the binary is live.
Furthermore, we must note that assuming a fixed binary orbit implies that the angular momentum is not properly conserved throughout the simulation. 

The main important conclusion is that fixing the orbit of the binary while evolving the gas distribution leads to an overestimate of the positive gravitational torque together with a more severe underestimate of the accretion torque onto the binary.
The absence of the accretion rate periodicity on the binary orbital period in the fixed orbit simulation has also very important implications since this is one of the properties that can characterize a binary signal in the observations.

We finally note that the study we performed was limited to circular, equal mass binaries. 
The only existing study of eccentric planets (i.e. binary mass ratios $\lesssim 10^{-3}$) embedded in gaseous discs that compare fixed and moving planets has been carried out by \cite{BitschKley2010}.
In particular, their Figure 24 shows that in radiative discs fixing the planet orbit leads to a higher critical eccentricity below which the planet is more prone to outward migration.  

\section*{Acknowledgments}

 We thank the anonymous referee for providing useful comments that improved the clarity of the paper.
We thank Daniel Price for providing the {\sc phantom} code for numerical simulations and acknowledge the use of {\sc splash} \citep{Price2007} for the rendering of the figures.
We thank Phil Hopkins for providing the {\sc gizmo} code for numerical simulations. 
AF and AS acknowledge financial support provided under the European Union’s H2020 ERC Consolidator Grant ``Binary Massive Black Hole Astrophysics" (B Massive, Grant Agreement: 818691).
AL acknowledges funding from MIUR under the grant PRIN 2017-MB8AEZ.
ZH acknowledges support from NSF grants AST-2006176 and 1715661 and NASA grant 80NSSC22K082.
This research was supported in part by the National Science Foundation under Grant No. NSF PHY-1748958.

\section*{Data Availability}

Hydrodynamic simulations used the {\sc gizmo} code. The input files for generating the MFM simulations will be shared on reasonable request to the corresponding author.



\bibliographystyle{mnras}
\bibliography{references.bib} 





\bsp	
\label{lastpage}
\end{document}